# The Evolutions of the Rich get Richer and the Fit get Richer Phenomena in Scholarly Networks: The Case of the Strategic Management Journal

**Guillermo Armando Ronda-Pupo**
http://orcid.org/0000-0002-9049-8249
Departamento de Administración
Universidad Católica del Norte, Chile
AvenidaAngamos. Nº 0610,
Antofagasta, Chile
1270709.
gronda@ucn.cl

Departamento de Turismo
Universidad de Holguín, Cuba
Avenida XX Aniversario, Piedra Blanca
Holguín, Cuba
80100

**Thong Pham**
Mathematical Statistics Team,
RIKEN Center for Advanced Intelligence Project,
Tokyo,
Japan
thong.pham@riken.jp

All authors designed the experiments, performed the experiments, analyzed the results, wrote and reviewed the manuscript.

**Correspondence to:** Guillermo Armando Ronda-Pupo. Departamento de Administración. Universidad Católica del Norte. AvenidaAngamos. Nº 0610. Antofagasta, Chile. ZIP: 1270709.
Email: gronda@ucn.cl
Phone: +56 2552355802.

## Abstract

*Understanding how a scientist develops new scientific collaborations or how their papers receive new citations is a major challenge in scientometrics. The approach being proposed simultaneously examines the growth processes of the co-authorship and citation networks by analyzing the evolutions of the rich get richer and the fit get richer phenomena. In particular, the preferential attachment function and author fitnesses, which govern the two phenomena, are estimated non-parametrically in each network. The approach is applied to the co-authorship and citation networks of the flagship journal of the strategic management scientific community, namely the Strategic Management Journal. The results suggest that the abovementioned phenomena have been consistently governing both temporal networks. The average of the attachment exponents in the co-authorship network is 0.30 while it is 0.29 in the citation network. This suggests that the rich get richer phenomenon has been weak in both networks. The right tails of the distributions of author fitness in both networks are heavy, which imply that the intrinsic scientific quality of each author has been playing a crucial role in getting new citations and new co-authorships. Since the total competitiveness in each temporal network is founded to be rising with time, it is getting harder to receive a new citation or to develop a new collaboration. Analyzing the average competency, it was found that on average, while the veterans tend to be more competent at developing new collaborations, the newcomers are likely better at acquiring new citations. Furthermore, the author fitness in both networks has been consistent with the history of the strategic management scientific community. This suggests that coupling node fitnesses throughout different networks might be a promising new direction in analyzing simultaneously multiple networks.*

# Introduction

In a community of scholars, authors desire their research to have the highest influence possible (Rupp, Thornton, Rogelberg, Olien, & Berka, 2014). At present, the attractiveness of an author is measured by metrics, traditionally by the number of papers he/she writes and the number of citations these publications receive (Pan & Fortunato, 2014). Indeed, the core principle of a citation metric is the assumption that when an article is cited by another scholar, it has had an impact on their research (Neophytou, 2014). When a scholar's research is highly cited, that scholar is known to do high-impact research (Judge, Weber, & Muller-Kahle, 2012).

In 1993, the *Strategic Management Journal* (SMJ) with the co-sponsorship of Wiley and the *Strategic Management Society* (SMS), launched the Dan and Mary Lou Schendel *Strategic Management Journal Best Paper Prize* (SMJ BPP) to recognize those researchers that have significantly contributed to the advancement of the discipline. The BPP is awarded to a SMJ paper which was published at a minimum of five years prior to the award citation itself (Strategic Management Society, 2014). One of the criteria for evaluating the recognition of the nominee paper is the number of citations the paper received from other papers from the discipline.

The academic corpus of different scientific disciplines and funding agencies need suitable indicators to accurately assess research performance. The race to find a universal accepted indicator fostered the emergence of a high amount of research metrics. Among the most popular of those metrics are citation counts, the Hirsh index (Hirsch, 2005) and its descendants and the G index (Egghe, 2013) and its derivations or the crown indicator (Moed, 2010; Waltman, van Eck, van Leeuwen, Visser, & van Raan, 2011). Albeit the contribution of those metrics to the advancement of research evaluation processes and policy formulation has been paramount, an important question remains unsolved: how to account for the heavy tail characteristic of citations and publications distribution?

The above-mentioned question is raised on the basis of results presented in previous studies that demonstrated that the science system is characterized by scale invariant properties (Katz, 1999, 2016b; van Raan, 2008). These properties also hold true at the level of scientific domains (Katz, 2016a; Ronda-Pupo & Katz, 2017), universities (van Raan, 2013) or cities as innovative systems (Bettencourt, Lobo, Helbing, Kuhnert, & West, 2007), suggesting the scale free behavior is present at different scales of the science system. The traditional scientometric indicators are not capable of capturing the properties of the scale invariance of such systems.

Likewise, the scientometric approach for the analysis of co-authorship networks has focused mainly on elucidating the important actors in the network structure using centrality measures, such as degree centrality (Türker & Çavuşoğlu, 2016), betweenness (Abbasi, Hossain, & Leydesdorff, 2012; Guns, Liu, & Mahbuba, 2011) or closeness (Biscaro & Giupponi, 2014). What is lacking is a measure of the attractiveness of an author in a complex temporal network, not only in terms of their connectedness with other actors but also in terms of their fitness. To assess the fittest actors of that network it is important to define if a temporal network is governed by the preferential attachment phenomenon and to characterize the dynamics of its growth.

The process of knowledge creation and its dissemination through scientific journals is governed by the proportionate effect (Gibrat, 1931) or cumulative advantage phenomenon (de Solla-Price, 1976), also called by the synonyms the *rich get richer* (Simon, 1955), the Mathew Effect (Merton, 1968, 1988) and preferential attachment (Barabási & Albert, 1999). All these synonymous terms describe a phenomenon in which a well-connected author in a collaborative or citation network has a higher probability to gain new collaborators or citations than a less connected author. Since this phenomenon is believed to be at work in diverse types of complex networks, it has attracted the attention of several scientific communities (Bianconi & Barabási, 2001; Caldarelli, 2007; Caldarelli, Capocci, De Los Rios, & Munoz, 2002; Newman, 2001a).

On the other hand, the *fit get richer* phenomenon describes a process in which the numbers of new citations or new collaborations an author receives is proportional to a number called fitness, regardless of whether he or she is well-connected (Bianconi & Barabási, 2001; Wang, Song, & Barabasi, 2013). For example, it is reasonable to assume that two similarly well-connected authors will gain new citations or new collaborations according to the qualities of their sciences. In such a situation, author fitness can be interpreted as a proxy for the intrinsic quality of the scientific contributions of an author to the advancement of a discipline.

The combination of the rich get richer and the fit get richer mechanism can describe various dynamic patterns in a temporal network. For example, as will be discussed in the Background Section, the rich get richer mechanism can explain the *first-mover advantage*, a frequently-assumed advantage that pioneers of a field enjoy over latecomers, while the fit get richer mechanism can explain why occasionally a latecomer surpasses the pioneers. In other words, the rich get richer and the fit get richer mechanisms can *simultaneously* describe both the concentration process that forms an established expert as well as the emerging process of a new rising star; two processes that arguably exist in all innovation fields.

The co-authorship network in documents published in the Strategic Management Journal has been analyzed to trace the formation and evolution of the international scientific community of the discipline (Ronda-Pupo & Guerras-Martín, 2010) and to study the inter-institutional collaboration networks of the field (Koseoglu, 2016). These previous papers focused mainly on determining the most influential countries and institutions according to their degree centrality in the structure of the collaboration network.

The aim of the present study is to extend the knowledge of the strategic management scientific community by simultaneously analyzing the rich get richer and the fit get richer phenomena in the temporal co-authorship and citation networks of documents published in SMJ to determine the most attractive authors of the discipline. Furthermore, the evolutions of the rich get richer and the fit get richer phenomena themselves will be explored. The PAFit (Pham, Sheridan, & Shimodaira, 2015, 2016), which is a statistical method for measuring the preferential attachment and author fitness in a temporal complex network, will be used. This method implements mathematical algorithms that accurately detect if the evolution of a scientific community network is governed by the rich get richer and the fit get richer phenomena. The method is more accurate than methods that estimate the two phenomena separately, such as Newman's method (Newman, 2001a) or Kong's method (Kong, Sarshar, & Roychowdhury, 2008).

The research questions for the present study are:

*RQ1*: *Has the preferential attachment process been characterizing the temporal co-authorship network and secondly the citation network of the papers published in the Strategic Management Journal?*

*RQ2*: *Similarly, has the rich get richer phenomenon also been governing the co-authorship and citation networks of the papers published in the Strategic Management Journal?*

*RQ3*: *Who are the most influential authors of the strategic management field according to their author fitness in the co-authorship and citation network in which they participate?*

All documents published in SMJ between its inception in 1980 through September, 2017 were analyzed. This interval is divided into four consecutive periods in order to measure the evolutions of preferential attachment and author fitness. The results are aimed at scholars, editors, business schools, practitioner and PhD students interested in the evolution of strategic management as an academic research discipline and also, to scholars and researchers of network science and complex systems.

## Background

**A brief commentary on the use of Network Science in scientometric studies**

The introduction of the notions of network science in scientometric studies dates back to the Derek de Solla Price study of "Networks of Scientific Papers" (de Solla-Price, 1965). This paper can be considered one of the founding papers of the present Scientometric discipline. In general, from 1975 through 2018, scholars in Information Science and Library science have published 318 studies that employs a network analysis approach as their research method. From 1991 through 2018 the journal Scientometrics accounts for the 27% of the overall scientific production (85 articles) of network analysis with scientometric purposes, showing a sustained increase of scientometric studies using network analysis in the past decade.

The most cited network science articles among the Scientometrics studies are Leydesdorff (2007), Otte and Rousseau (2016) and van Eck and Waltman (2010). The most cited network science articles in the Scientometrics literature are Newman (2001b, 2001c), Barabási et al. (2002), Freeman (1978), Borgatti, Everett, and Freeman (2002) and Wasserman and Faust (1999).

Three main lines of research can be observed in the network science scientific production in Scientometric studies: to study scientific collaboration networks (Abbasi et al., 2012; Barabási et al., 2002; Biscaro & Giupponi, 2014; Guns et al., 2011; Koseoglu, 2016; Newman, 2001b, 2001c; Ronda-Pupo & Guerras-Martín, 2010, 2013; Türker & Çavuşoğlu, 2016), to unveil and present knowledge maps of scientific fields (Law & Whittaker, 1992; Otte & Rousseau, 2016; Ronda-Pupo, 2015; van Eck & Waltman, 2010) and a new line introduced recently by Batagelj, Ferligoj, and Squazzoni (2017) to detect the emergence of scientific fields using network analysis.

Previous papers used centrality measures to determine the most important nodes within the co-authorship collaboration networks or in co-word network structures (Leydesdorff, 2007; Newman, 2001b). Studies that measure the preferential attachment and author fitness in consecutive periods of two scholarly temporal networks at the same time is lacking in the literature. Such a study has two merits: a) it can reveal the evolutions of the rich get richer and fit get richer phenomena themselves and b) the two phenomena can be compared in two different networks to reveal network-specific patterns. The present study is aimed at filling this gap.

*Preferential Attachment*

Preferential attachment (PA) is a stochastic process that has been proposed to explain certain topological features characteristic of complex networks from diverse domains (Pham et al., 2015). The term, PA, was coined by Barabási and Albert (1999) in network science. It has its roots in the Gibrat's law or rule of proportionate growth or the law of Proportionate Effect (Gibrat, 1931). Gibrat stated that the proportional rate of growth of a firm is independent of its absolute size. Merton (1968, 1988) called the Gibrat's law success-breeds-success phenomenon. It has also been called by the synonyms the *rich get richer* (Simon, 1955), *cumulative advantage* (de Solla-Price, 1976) or the *Matthew Effect* (Merton, 1968) in the scientometric literature.

The studies of scaling behavior in bibliometric studies dates back to the studies of citation networks carried out by Alfred Lotka (Lotka, 1926) and complemented later by Dereck de Solla-Price (de Solla-Price, 1965). Naranan (1971) introduced the power law approach to study the Bradford's law in scientific journals. These studies were the theoretical foundations of the Egghe (2005) *Lotkaian informetric*. The Lotkaian approach has been focused mainly on characterizing distributions which satisfy Price's Law and consequences for the Laws of Zipf and Mandelbrot (Egghe & Rousseau, 1986).

In mathematical terms, the PA mechanism states that an author with $k$ citations will get a new citation with probability proportional to $A_k$, the PA function. The rich get richer phenomenon exists if $A_k$ is an increasing function on average, since in that case a highly connected node will acquire more edges than a lowly-connected node. In bibliometrics language, this means that a highly cited author has more probability of being cited again than a non-cited author. This process fosters the emergence of a disproportionate cumulative recognition of an elite of authors and leads to the characteristic heavy tail distributions.

The rich get richer phenomenon also leads to an explanation of the first-mover advantage, an influential concept in theories of strategic management. Originally this concept describes the advantage firms enjoy when they are pioneers in a new market when compared to late-coming firms (Frynas, Mellahi, & Pigman, 2006). In the context of citation networks, it is the advantage to accrue more citations for authors who enter a scientific subject matter in its infancy rather than latecomers (Newman, 2009). The first-mover advantage can be explained by the rich get richer mechanism: an early node will have more time to accumulate links, and thus reinforces its advantage through the rich-get-richer mechanism. More specifically, the age of a node is positively correlated with its total number of citations when the rich get richer phenomenon alone is at play.

The properties of a PA function can often be summarized by assuming the log-linear form $k^\alpha$ and measuring the attachment exponent $\alpha$. The attachment exponent is important since it reveals many properties of the PA function. For example, the rich get richer phenomenon exists if $\alpha > 0$ and not if $\alpha < 0$. The attachment exponent also reveals properties of the temporal network in the long term: when $\alpha < 1.0$, the sub linear case, the degree distribution of the network is a stretched exponential, while in the super linear case of $\alpha > 1.0$, one node will eventually get all incoming new links (Krapivsky & Redner, 2001). This situation has been called the *winner take all* effect.

*Author Fitness*

Node fitness captures individual differences between nodes which have the same number of links-differences and are ignored in the PA mechanism. In network terminologies, the probability a node $v_i$ receives a new link is proportional to its fitness $\eta_i$. This means that in a citation network, two authors with the same number of citations would get cited proportionally to their respective fitness, which can be interpreted as the qualities of their scientific contributions.

Author fitness explains why late-comers can surpass the first-movers, a situation that could not happen under the rich get richer mechanism. In citation networks, it has been observed that some late-comers acquire more citations than the first-movers (Newman, 2009). Such deviations from the first-mover advantage can be explained by author fitness: if an author has high enough fitness, then even if the author is late to the game, the high fitness would help the author to surpass the first-movers. This way, fitness becomes a competitive advantage of authors in a citation network.

*Combining Preferential Attachment and Author fitness*

In the present study, both the concepts of PA as rich get richer and of author fitness as fit get richer in the co-authorship and citation temporal networks of the strategic management scientific community are studied simultaneously using the flagship journal of the discipline, the SMJ. The simultaneous combination of the rich get richer and the fit get richer phenomenon reveals the dynamic of the temporal networks: for example, it explains at the same time the first-mover advantage and deviations from such advantage.

## Methods

**The experiment**

*Time frame*

A universal method to select study time frames to analyze the dynamics of the evolution of a scientific community does not exist (Ronda-Pupo & Guerras-Martín, 2013). Previous bibliometric papers in this area included different time frames for their study. For example, Ramos-Rodríguez and Ruíz-Navarro (2004), as well as Nerur, Rasheed, and Natarajan (2008), analyzed 21 years of scientific production published in the *Strategic Management Journal* (1980 – 2000), segmented into three stages. Furrer, Thomas, and Goussevskaia (2008) studied 26 years of documents published in the *Academy of Management Journal (AMJ), Academy of Management Review (AMR), Administrative Science Quarterly (ASQ),* and *SMJ* (1980 – 2005) segmented into five stages. Nag, Hambrick, and Chen (2007) used for their study a 21-year time frame of articles on strategic management in the *SMJ, AMJ, AMR,* and *ASQ* (1980 – 2000), segmented into five stages. All these studies focused their attention on finding the dynamics of the intellectual structure of the discipline using bibliometric methods.

An analysis of temporal co-authorship networks has been carried out by Ronda-Pupo and Guerras-Martín (2010) that analyzed 30 years of the international co-authorship network in documents published in SMJ segmented into three segments of 10 years each (1980-2009). Recently, Koseoglu (2016) analyzed 34 years of documents published in SMJ, segmented into 5 time spans (1980-2014). In the present study, the time frame was extended to 2017. The data is segmented into four time-spans. The first time span covers from 1980 through 1989, the second interval from 1990 through 1999, the third 2000-2009 and the last time span covers from 2010 through September, 2017.

*The Data*

The data for the study consists of all documents published in the *Strategic Management Journal* since its inception in 1980 through September, 2017. SMJ is used to study the temporal citation and co-authorship networks of the strategic management scientific community because it is generally considered to be the flagship journal of the strategic management discipline (Azar & Brock, 2008; Guerras-Martin & Ronda-Pupo, 2013; Tahai & Meyer, 1999).

There is a twofold preparation of datasets to determine the PA function and author fitness: 1) the co-authorship temporal network, and 2) the citation temporal network. Each of these steps is described below.

*The co-authorship temporal network*

Co-authorship is traditionally used to analyze collaboration among scholars in any scientific discipline. This approach assumes that if an author signs a paper with another, it involves a collaboration to create and disseminate knowledge. To prepare the co-authorship network we used the signing authors of each SMJ document. When an article has one author, that author is included in the network as an isolated node if he or she is not already appeared. The result is an undirected co-authorship network including 2704 nodes (authors) and 8262 edges (links). The academic birth time of an author in the co-authorship network is considered to be the date of the first paper they had published in SMJ. The degree distribution over time in the co-authorship network is shown in Figure 1.

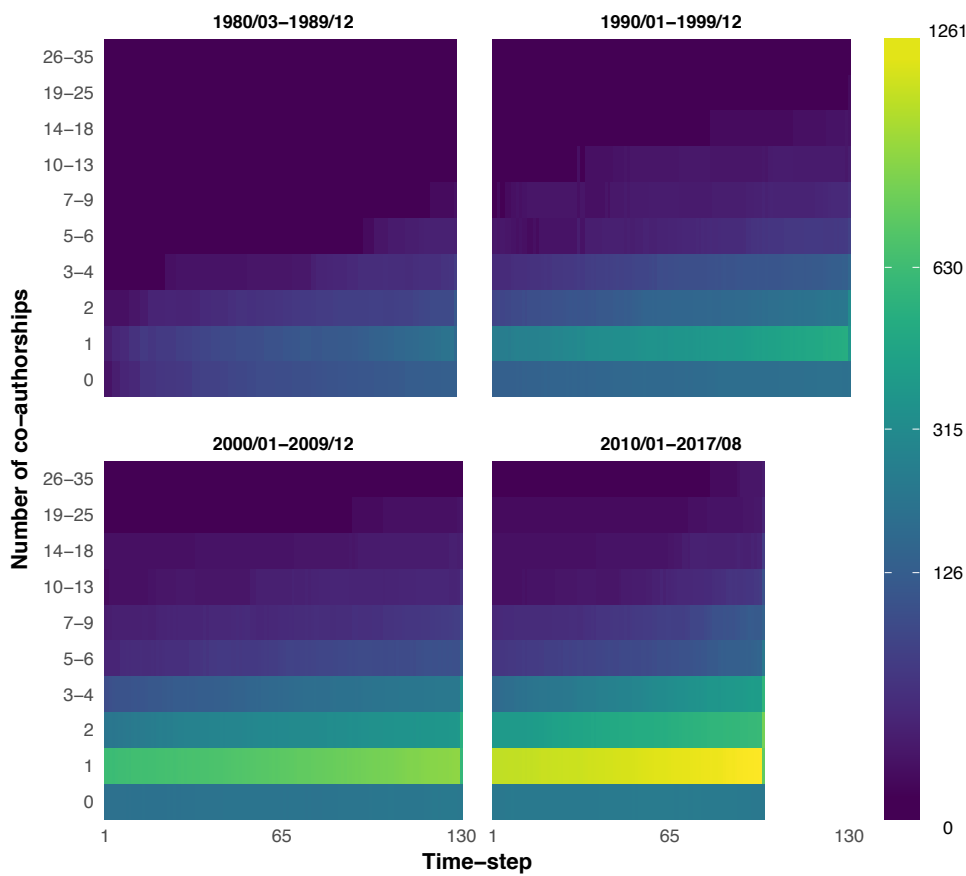

**Fig.** 1: Number of authors with $k$ co-authorships. The temporal co-authorship network is divided into four periods. The time resolution is monthly. In each period, the first time-step corresponds to the first month in that period. Each cell shows the number of authors with that number of co-authorships.

*The citation temporal network*

The citation network consists of all citations in the documents published in SMJ in each time span analyzed. To build the temporal citation network, the bibliometrics Package BIBEXEL (Persson & Danell, 2009) was used. The academic birth time of an author in the citation network is considered to be the date of the first citation he/she received in SMJ.

The result is a directed citation network including 2077 nodes (cited authors) and 73585 edges. Only citations to authors that were published in SMJ were included. The preparation of both networks required disambiguation of 108 author's names. For example, the author Igor Ansoff, appeared as Ansoff, HI; Ansoff, H.I.; Ansoff, H; Ansoff, I; Ansoff, H.I. The author Luis Gómez-Mejia, appeared as Gomezmejia, L or Gómez Mejia, L. This process was carried out manually using Excel. The degree distribution over time in the citation network is shown in Figure 2.

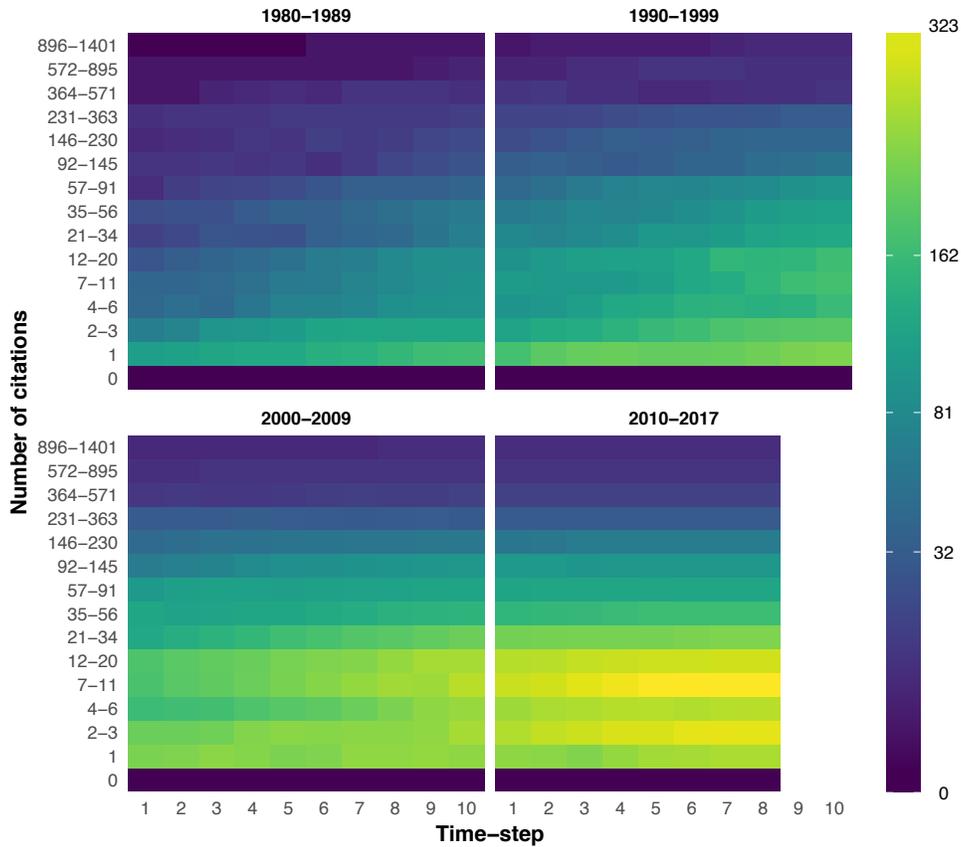

**Fig.** 2: Number of authors with *k* citations. The temporal citation network is divided into four periods. The time resolution is yearly. In each period, the first time-step corresponds to the first year in that period. Each cell shows the number of authors with that number of citations.

**Procedure to calculate the preferential attachment and author fitness**

Assuming both the rich get richer and the fit get richer phenomena are at play, the probability $P_i(t)$ that a node $v_i$ gets a new link at time t is:

$$P_i(t) \propto A_{k_i(t)} \times \eta_i, \qquad (1)$$

where $k_i(t)$ is the degree of node $v_i$ at time $t$, $A_k$ is the PA function and $\eta_i$ is the fitness of node $v_i$. In bibliometric terms, $P_i(t)$ can be interpreted as the total attractiveness of an author $v_i$ at time $t$, which is the product of the author's intrinsic scientific quality (the fitness $\eta_i$) and the author's well-connectedness at that time (the PA value $A_{k_i(t)}$). $A_k$ and $\eta_i$ are normalized so that $A_0 = 1$ and the mean of all $\eta_i$'s is 1.

Based on Equation (1), the log-likelihood of the observed temporal network can be explicitly written and the PA function $A_k$ and author fitness $\eta_i$ can be estimated by maximizing the log-likelihood function with suitable regularization terms (Pham et al., 2016). After obtaining the estimated value of $A_k$, the log-linear functional form $A_k = k^\alpha$ can be fitted to the estimated $A_k$ to find the attachment exponent $\alpha$. To perform those tasks, the R package PAFit (Pham, Sheridan, & Shimodaira, 2017) was used.

After estimating the PA function $A_k$ and node fitness $\eta_i$, one can measure how hard it is to get a new edge at time $t$ by calculating the *total competitiveness* at time $t$, denoted $S(t)$, which is defined as the sum of the un-normalized probabilities $P_i(t)$ of all nodes that existed at time $t$:

$$S(t) = \sum_{i:\, v_i \text{ existed at time } t} A_{k_i(t)} \times \eta_i. \qquad (2)$$

The larger $S(t)$ is, the harder it is to get a new edge at time $t$. To see this, imagine an "average" node with degree 0 and fitness 1 at time $t$ (node fitnesses are normalized so that their mean is 1). The attractiveness of such a node, calculated by Equation 1, is equal to 1. If $S(t)$ is increasing, it will be harder for this average node to get a new edge.

The increase in competition might be due to an increase in the number of new players that enter the field. To separate such an effect, one might want to consider the *average competitiveness* $\overline{S(t)}$, which is defined as:

$$\overline{S(t)} = \frac{S(t)}{N(t)}, \qquad (3)$$

where $N(t)$ is the number of nodes that existed at time $t$.

Another quantity of interest is the *average competency* $C(t)$, defined as the average of node fitnesses of all nodes that existed at time $t$:

$$C(t) = \frac{1}{N(t)} \sum_{i:\, v_i \text{ existed at time } t} \eta_i. \qquad (4)$$

$C(t)$ reveals the trend of the competency of the authors: an increasing $C(t)$ would mean that on average, the new authors tend to have higher fitness than the old authors, while a decreasing $C(t)$ would mean that on average, the old authors are likely more competent than the new authors.

## Results

Figure 3 shows the estimated PA functions $A_k$ and the attachment exponents $\alpha$ of the co-authorship network using PAFit in four stages. The results show that the rich get richer phenomenon is at play, since in every stage the PA function $A_k$ is increasing in $k$. The PA functions did not change much between each of the four time-stages, which suggests that the rich get richer phenomenon is time-stable. All the attachment exponents are less than 0.50, which indicates that the rich get richer phenomenon is rather weak. This means that when freshman authors of the SMJ co-authorship network seek out collaborations, they look for already well-connected or senior authors but do not only rely on that well-connectedness.

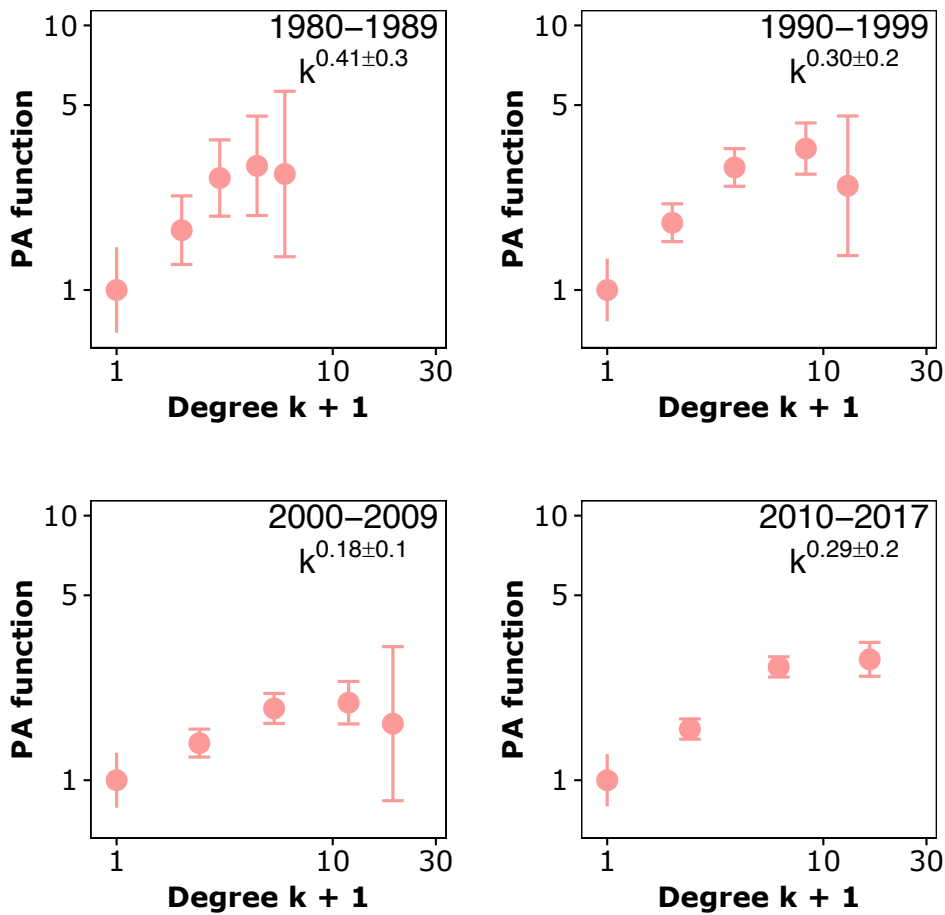

**Fig.** 3: Estimation of the PA functions and the attachment exponents in four-time intervals of the co-authorship network using PAFit. The bar at each estimated point is the estimated two-sigma confidence interval. The estimated attachment exponent in each time interval together with its standard deviation are shown in the upper left corner of the corresponding panel.

Figure 4 shows the estimated PA functions $A_k$ and the attachment exponents $\alpha$ of the citation network using PAFit in four stages. As in the co-authorship network, the rich get richer phenomenon is present. The attachment exponents are sublinear and are generally on the same order as those in the co-authorship network in the four stages analyzed. The PA functions also did not change much through the four stages, which suggest that the rich get richer phenomenon is time-stable. The attachment exponent $\alpha$, however, slightly decreased with respect to time. This pattern could mean that authors rely less and less on the well-connectedness of a paper in choosing which papers to cite.

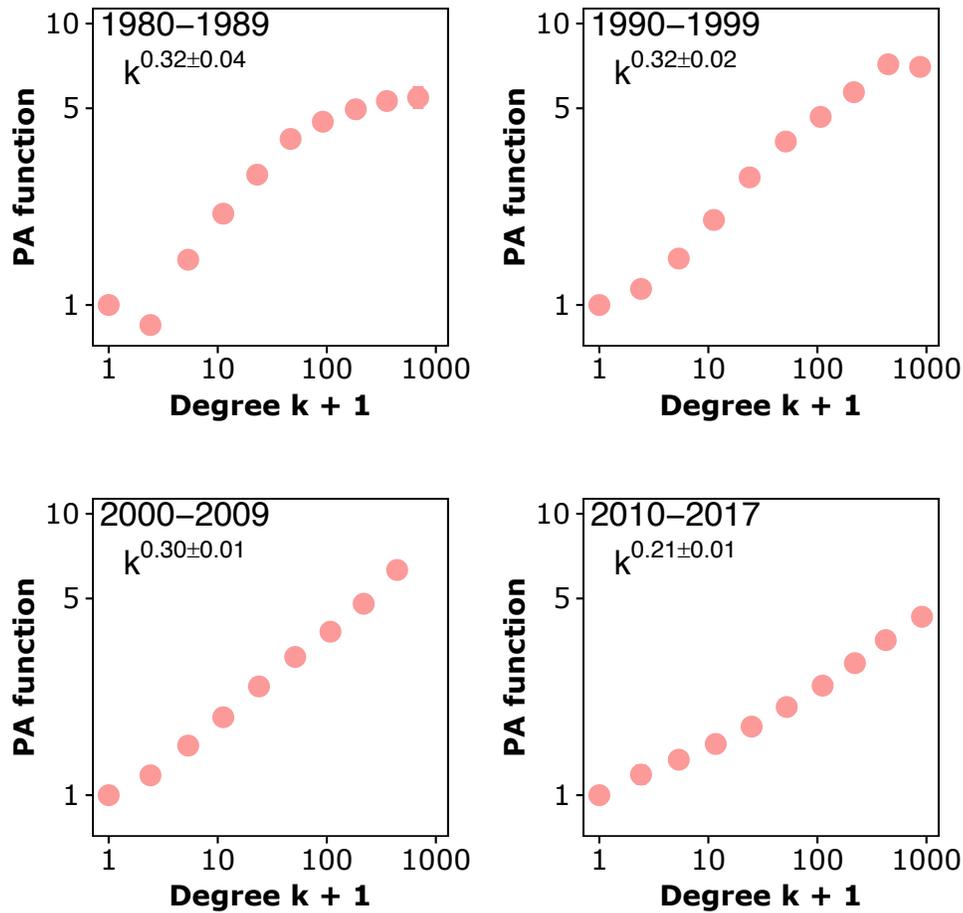

**Fig.** 4: Estimation of the PA functions and the attachment exponents in four intervals of the citation network using PAFit. The two-sigma confidence intervals are of the same size as the estimated points. The estimated attachment exponent in each time interval together with its standard deviation are shown in the upper left corner of the corresponding panel.

Figure 5 shows the histograms of author fitness of the co-authorship network in four intervals. For all these intervals, the histograms did not concentrate around 1 but have long right tails, which suggest the existence of the fit get richer phenomenon. The increase of the frequencies (bar heights) with time reflects the fact that more and more new authors entering the system. The range of author fitness in each period is increasing with time too: the maximum fitness value in each period is 4.24, 9.31, 22.53 and 26.92. This suggests the co-authorship network becomes increasingly competitive through time, a fact that will be explored later.

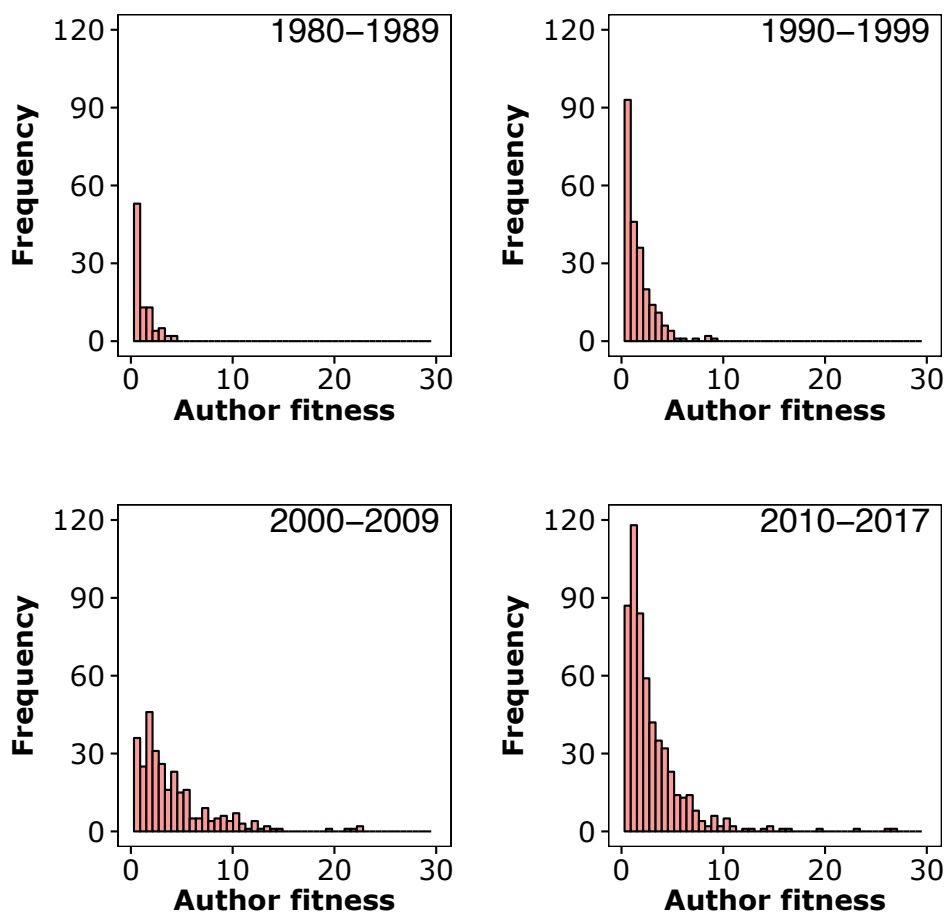

**Fig**. 5: Histogram of author fitness estimated by PAFit in four intervals of the co-authorship network. The heavy tails of these distributions imply the existence of the fit get richer phenomenon.

The histograms of author fitness of the citation network in the four intervals (Figure 6) reveals the same pattern that was observed in the co-author network: none of the histograms concentrate around 1, which clearly suggests the existence of the fit get richer phenomenon in the citation network from 1980 to 2017. Compared to the histograms in Figure 3, the right tails of those histograms in Figure 4 are shorter, which suggests the fit get richer phenomenon is weaker in citation network than in co-authorship network. Furthermore, the maximum fitness value in each period is 8.69, 14.08, 11.26 and 8.40, which suggests that competition in the citation network became fiercer and fiercer at least until the 1990-1999 period.

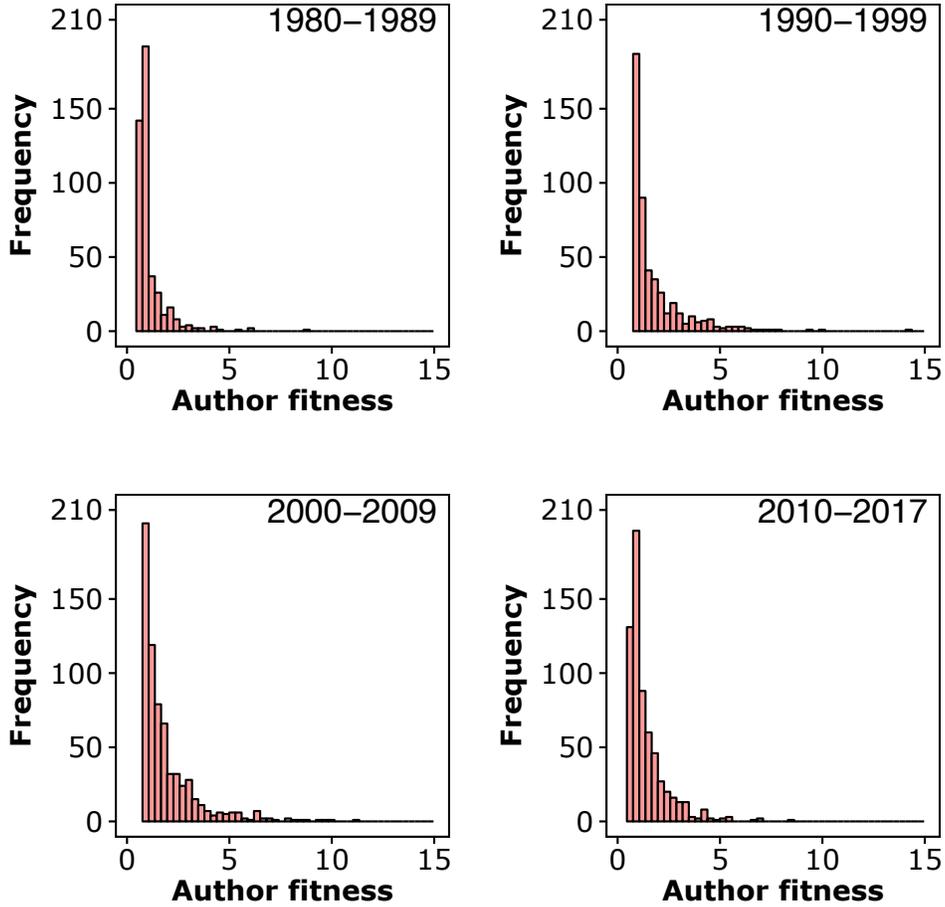

**Fig**. 6: Histogram of author fitness estimated by PAFit in four intervals of the citation network. The heavy tails of these distributions imply the existence of the fit get richer phenomenon.

Next, how difficult for a node to get a new edge is analyzed by calculating the total competitiveness $S(t)$ and the average competitiveness $\overline{S(t)}$ in the co-authorship and citation networks. Figures 7 and 8 show $S(t)$ and $\overline{S(t)}$ in the co-authorship network and the citation network, respectively. They have been normalized so that $S(t_0) = \overline{S(t_o)} = 1$ in both networks, where $t_0$ is the initial time of the network. Both $S(t)$ and $\overline{S(t)}$ increased with time in the two networks, which implies that the competition for a new edge did become fiercer and fiercer. In Figure 7, the total competitiveness $S(t)$ of the co-authorship network rose nearly 300 times from its initial value. Almost all of this increase, however, is due to the influx of new authors, since $\overline{S(t)}$ rose only 1.5 times, i.e., the influx of new authors accounts for 99.5 % of the increase of $S(t)$ in the co-authorship network. On the other hand, in Figure 8, $S(t)$ and $\overline{S(t)}$ of the citation network rose 10 times and 2 times, respectively. This means that the influx of new authors only explains about 80 % of the increase in $S(t)$ in the citation network.

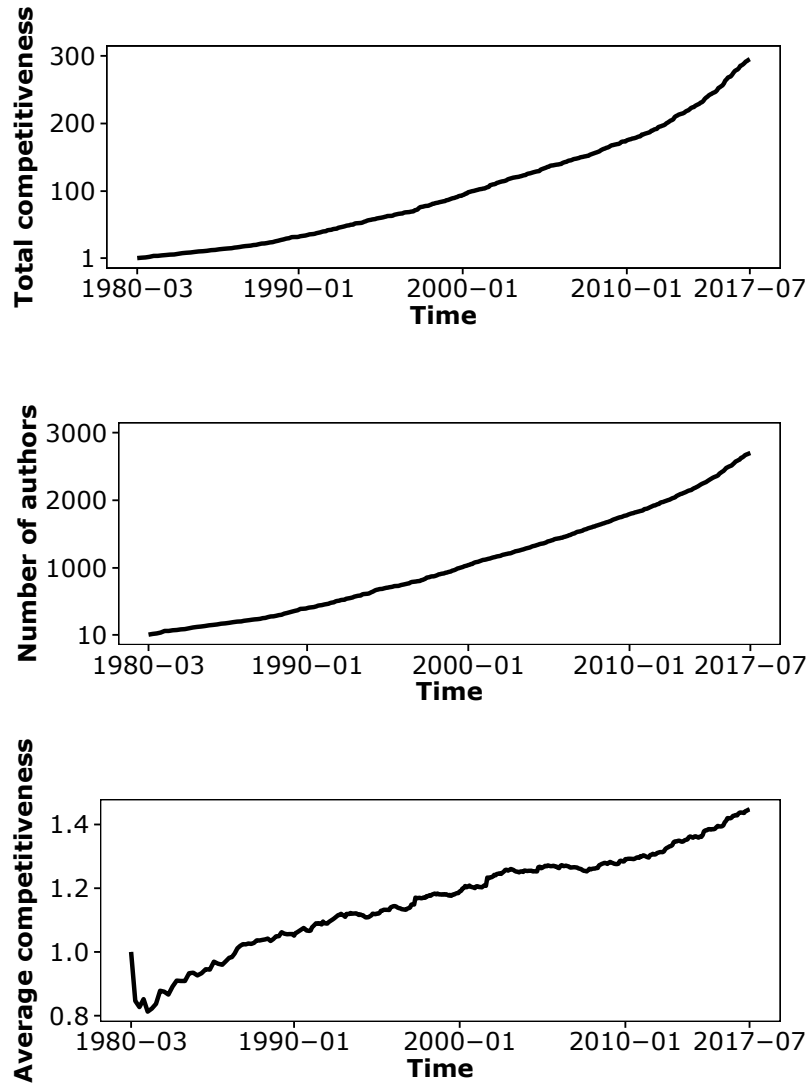

**Fig**. 7: Total competitiveness $S(t)$, number of authors, and average competitiveness $\overline{S(t)}$ in the co-authorship network. An increase in $S(t)$ implies that it is getting harder to get a new co-authorship. Most of this increase, however, is due to the increase of the number of authors in the network, since the average competitiveness $\overline{S(t)}$ did not rise significantly.

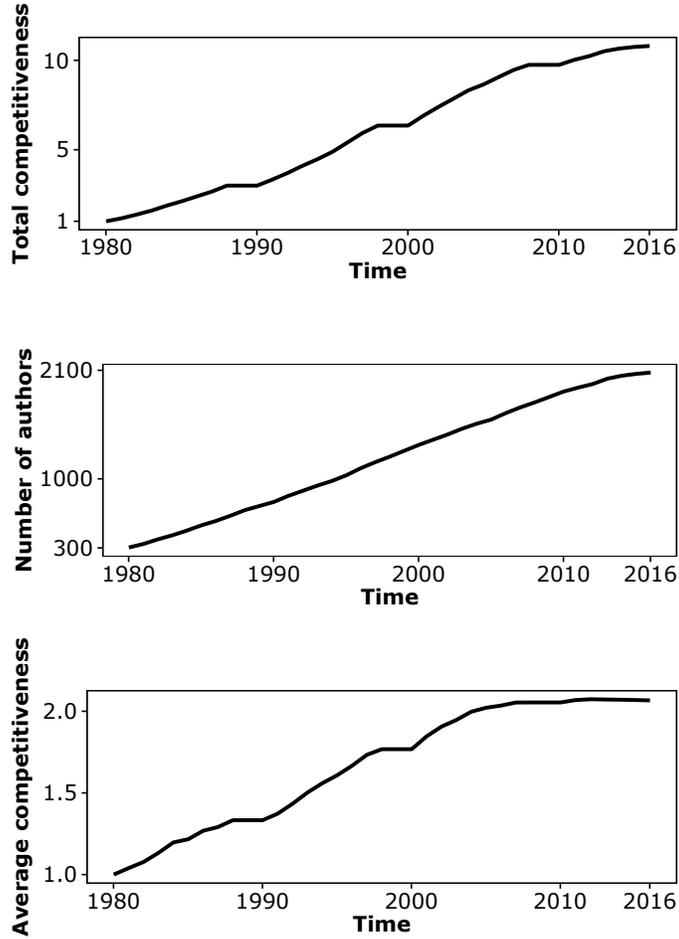

**Fig**. 8: Total competitiveness $S(t)$, number of authors, and average competitiveness $\overline{S(t)}$ in the citation network. An increase in $S(t)$ implies that it is getting harder to get a new citation. Even if the increase of the number of authors is taken into account, the average competitiveness $\overline{S(t)}$ still rose significantly. $\overline{S(t)}$, however, remained relatively flat from 2008 to 2017, which suggests that on average, the competition for new citations did not rise in this period.

The remaining portion of the increase of $S(t)$ in the citation network is partly due to an increase in the competency of authors in the citation network. In Figure 9, the average competency $C(t)$ is shown during the growth process of each network. It was also normalized so that $C(t_0) = 1$ in both networks, where $t_0$ is the initial time of the network. In the co-authorship network, $C(t)$ is decreasing overall, which implies that on average the veterans are likely more competent in getting new co-authorships than the newcomers. The situation in the citation network, however, is reversed: $C(t)$ is increasing overall, which means that on average new authors tend to be more competent in getting new citations than the veterans.

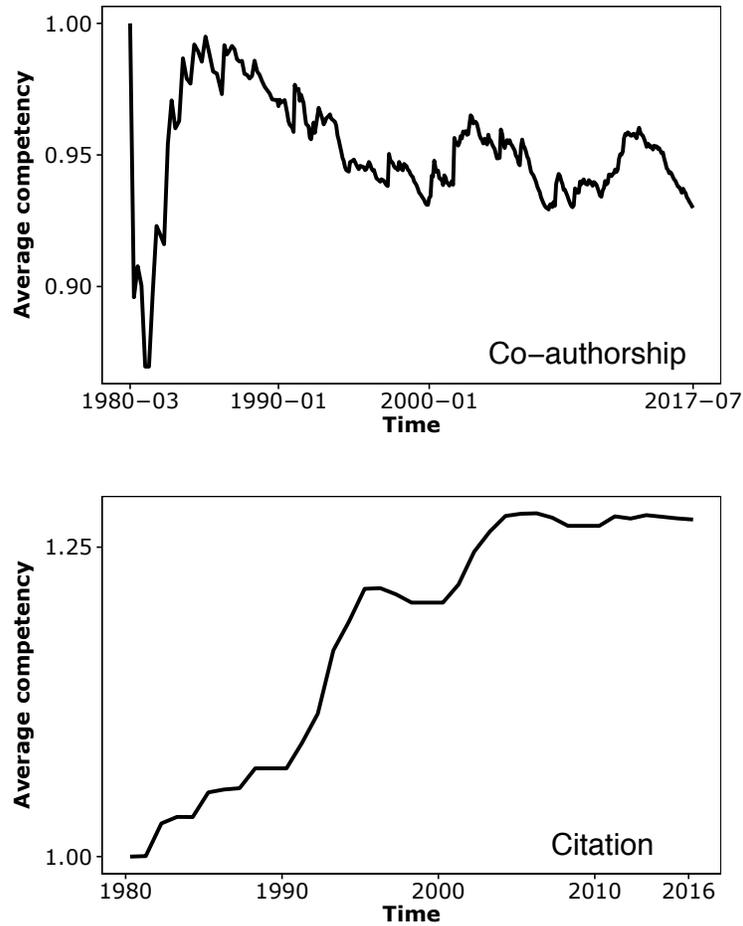

**Fig**. 9: Average competency $C(t)$ in the co-authorship and citation networks. A decreasing $C(t)$ in the co-authorship network implies that on average, the veterans tend to be better at getting new co-authorships than the newcomers. On the other hand, in the citation network, an increasing $C(t)$ implies that on average, the newcomers are likely more competent in getting new citations than the veterans. The relative flatness of $C(t)$ in the citation network between 2008 and 2017 suggests that the new authors in this period did not propose new approaches that could raise their competencies.

The average competency in the citation network, however, is relatively flat in the period from 2008 to 2017. This coincides with the relative flatness of $\overline{S(t)}$ at the same period (Figure 8). This suggests that the new authors in this period did not propose approaches that were novel enough to increase their competencies comparing with those of the old authors. This in turn leads to the competition for new citations among all authors did not rise on average.

Table 1 shows the results of the estimation of author fitness of the co-authorship and the citation temporal networks using PAFit. The most attractive authors in the co-authorship network were Michael Hitt and Will Mitchell. These authors were attractive in three out the four intervals analyzed. Richard Bettis and David Ketchen were among the most attractive authors in 50% of the time spans.

When results from both temporal networks, co-authorship and citation, were considered simultaneously, the most attractive authors of the strategic management scientific community were Richard Bettis, Catherine Helfat and Birger Wenerfelt.

All top-ten authors of the 1980-1989 decade introduced important subject matters to the nascent strategic management as an academic research discipline. These authors introduced relevant theoretical approaches such as: the *resource-based view* of the firm, the *dynamic capabilities*, *upper echelons*, *competitive advantage*, *alliances & networks* and *agency theory*. All these topics became hot research subject matters and received significant attention by the strategic management scientific community. The contributions of these authors to the advancement of the theoretical corpus of the discipline are highlighted by the following.

The *resource-based view* of the firm, developed by Wernerfelt (1984) was awarded with the 1994 SMJ Best Paper Prize. The academic interest in this line of research grew and was developed by other authors in the following decades. The most notable were Peteraf (1993), whose contribution received the 1999 SMJ BPP and recently Barney (2016). These authors, Margarette Peteraff and Jay Barney appear among the top fittest authors in the 1990-99 stage.

*Alliances &networks* is another important line of research. It was introduced by Kogut (1988) who was awarded with the 1998 SMJ Best Paper Prize. This theoretical approach was developed later by scholars such as Ring and van de Ven (1992), whose contributions were awarded with 2008 SMJ BPP, and later Gulati (1998) continued to develop the *alliances & network* approach in the 2000-2009 decade. His contribution received the 2014 SMJ Best Paper Prize. The author Ranjai Gulati is the fittest author of the 1990-99 stage.

The *competitive advantage* approach presented by Porter (1980) was also among the most important lines of research of the strategic management discipline. This author, Michael Porter, became the most cited author in the SMJ history. Henderson and Cockburn (1994) made important contributions to the advancement of this line of research and thus received the 2010 SMJ BPP.

Hambrick (1981), who is one of the most productive authors in the strategic management discipline developed the research line *upper echelons*. He has published 22 papers in SMJ. *Corporate strategy*, first introduced by Igor Ansoff in the 1960s, was developed by Richard Bettis and among others in the time frames analyzed in the present study.

The approach *dynamic capabilities,* introduced by Teece, Pisano, and Shuen (1997) is also among the most important research lines of the strategic management discipline. Their paper was awarded with 2003 SMJ BPP. This subject matter was complemented by contributions of Eisenhardt and Martin (2000). That contribution was worthy enough to obtain the 2007 SMJ BPP. Winter (2003) also received the 2009 SMJ BPP for his contribution to this topic.

**Table** 1. Ranks according to co-authorship and citation author fitness in each time interval.

| Time span | Co-authorship network | | Citation Network | |
|---|---|---|---|---|
| | Author | Fitness | Author | Fitness |
| 1980-1989 | Thomas H | 4.736 | Dierickx I | 8.185 |
| | Wernerfelt B | 4.718 | Hambrick D | 5.827 |
| | Bettis R | 4.116 | Wernerfelt B | 5.651 |
| | Montgomery C | 3.951 | Eisenhardt K | 5.151 |
| | Kim W | 3.610 | Kogut B | 4.461 |

|           | Hitt M        | 3.599  | Nelson R     | 4.201 |
|           | Robinson R    | 3.578  | Porter M     | 4.052 |
|           | Bracker J     | 3.496  | Teece D      | 3.948 |
|           | Macmillan I   | 3.146  | Miller D     | 3.486 |
|           | Pearce J      | 3.096  | Williamson O | 3.396 |
| 1990-1999 | Mitchell W    | 9.305  | Gulati R     | 8.088 |
|           | Lubatkin M    | 8.561  | Dyer J       | 5.926 |
|           | Smith K       | 8.250  | Barney J     | 4.620 |
|           | Hitt M        | 7.415  | Stuart T     | 4.256 |
|           | Hoskisson R   | 5.917  | Coff R       | 4.238 |
|           | Wright P      | 5.349  | Kogut B      | 4.129 |
|           | Daily C       | 5.088  | Peteraf M    | 4.102 |
|           | Miller A      | 4.799  | Baum J       | 4.095 |
|           | Dalton D      | 4.715  | Levinthal D  | 4.046 |
|           | Lane P        | 4.710  | Capron L     | 4.026 |
| 2000-2009 | Ketchen D     | 20.412 | Hoetker G    | 8.203 |
|           | Hitt M        | 21.294 | Rothaermel F | 7.741 |
|           | Keil T        | 20.538 | Zollo M      | 7.354 |
|           | Hult G        | 20.297 | Helfat C     | 7.301 |
|           | Li J          | 18.730 | Adner R      | 6.319 |
|           | Peng M        | 13.633 | Sirmon D     | 6.318 |
|           | Mitchell W    | 13.125 | Rosenkopf L  | 6.151 |
|           | De Sarbo W    | 12.940 | Carpenter M  | 6.027 |
|           | Semadeni M    | 12.355 | Lavie D      | 6.025 |
|           | Beamish P     | 12.104 | Villalonga B | 5.494 |
| 2010-2017 | Mitchell W    | 26.507 | Campbell B   | 5.387 |
|           | Bettis R      | 25.623 | Chatain O    | 5.082 |
|           | Gambardella A | 22.907 | Foss N       | 5.060 |
|           | Helfat C      | 19.336 | Bettis R     | 4.925 |
|           | Bell R        | 16.087 | Bingham C    | 4.031 |
|           | Agarwal R     | 15.390 | Crossland C  | 3.938 |
|           | Reuer J       | 14.637 | Quigley T    | 3.895 |
|           | Judge W       | 14.301 | Eggers J     | 3.838 |
|           | Miller D      | 13.975 | Connelly B   | 3.744 |
|           | Ketchen D     | 12.792 | Powell T     | 3.572 |

## Discussion and Conclusion

The results make evident that the evolution of the complex temporal co-authorship and citation networks of the strategic management scientific community are governed jointly by the rich get richer and the fit get richer phenomena. The rich get richer phenomenon is weak in both citation and co-author networks. This means that through nearly four decades, the intrinsic scientific quality of an author was the most important factor for that author to get new collaborations or new citations.

The increase of total competitiveness $S(t)$ in both networks shows that it is getting harder to get a new co-authorship or a new citation. While an increase in the number of authors contributes to the increase of $S(t)$ in both networks, there is another reason for the increase of $S(t)$ in the citation network: it is due to the increase in the competency of the authors. On average, while the veterans are likely more competent at getting new co-authorships than the newcomers, the newcomers tend to be better at getting new citations.

A correlation of the most attractive authors of the discipline, the research lines they contribute to and the relevance of these subject matters is confirmed in the history of the Strategic Management Journal Best Paper Prize. This correlation highlights that assessing the values of author fitness and PA within scientific domains or its research subject matters is an accurate method to predict the future relevant lines of research or important authors within the field.

At the same time, it is confirmed that the First Mover Advantage is enhanced by the rich get richer phenomena in a temporal complex network. Pioneer authors of a research line accumulate advantage with respect to latecomers. New authors cite pioneers proportional to their attractiveness leading to a disproportionate growth of their recognition fostering the emergence of a scale free or heavy tail distribution as found by de Solla-Price (1965).

The results show the efficacy of PAFit as an accurate method to assess PA and author fitness in complex temporal networks in scientific disciplines. It is possible to characterize the rich get richer and the fit get richer phenomena simultaneously. It is also possible to find authors that might be influential in the citation network because their fitnesses enhance their attractiveness despite not being among the most prolific in the co-authorship network.

The research in this subject matter could be enhanced with new research questions such as: Is it possible to estimate a co-authorship-fitness and a citation-fitness of the same author under the constrain that the two fitness values are the same? Do the properties of the rich get richer phenomenon observed in citation and co-authorship networks of the SMJ generalize to other citation and co-authorship networks? In particular, is the rich get richer phenomenon always the weaker phenomenon in those two networks? If that is the case, is using a unified PA function for multiple citation networks or multiple co-authorship networks better than using different PA functions? Is it possible to assess the fitness at the level of countries or institutions? Is it possible to elucidate the fitness of specific research topics of a given scientific field?

**Funding information**

This study was financed by FONDECYT, Chile, grant number 1180200 to Professor Guillermo Armando Ronda-Pupo.

**Acknowledgements**

The authors thank the two reviewers for interesting suggestions on a previous version of the manuscritp. We also express our gratitude to Professors Luis Ángel Guerras-Martin and Anoop Madok for interesting comments on the links between fittest authors with their lines of research.